\documentclass[
    ,final            
  ,sort&compress                 
  ]
  {aipproc}

\layoutstyle{6x9}

\begin{document}

\def\degr{\hbox{$^\circ$}}
\def\arcmin{\hbox{$^\prime$}}
\def\arcsec{\hbox{$^{\prime\prime}$}}
\def\la{\mathrel{\mathchoice {\vcenter{\offinterlineskip\halign{\hfil
$\displaystyle##$\hfil\cr<\cr\sim\cr}}}
{\vcenter{\offinterlineskip\halign{\hfil$\textstyle##$\hfil\cr
<\cr\sim\cr}}}
{\vcenter{\offinterlineskip\halign{\hfil$\scriptstyle##$\hfil\cr
<\cr\sim\cr}}}
{\vcenter{\offinterlineskip\halign{\hfil$\scriptscriptstyle##$\hfil\cr
<\cr\sim\cr}}}}}
\def\ga{\mathrel{\mathchoice {\vcenter{\offinterlineskip\halign{\hfil
$\displaystyle##$\hfil\cr>\cr\sim\cr}}}
{\vcenter{\offinterlineskip\halign{\hfil$\textstyle##$\hfil\cr
>\cr\sim\cr}}}
{\vcenter{\offinterlineskip\halign{\hfil$\scriptstyle##$\hfil\cr
>\cr\sim\cr}}}
{\vcenter{\offinterlineskip\halign{\hfil$\scriptscriptstyle##$\hfil\cr
>\cr\sim\cr}}}}}

\title{Nonthermal X-radiation  of  \\ SNR RX~J1713.7$-$3946: \\
The Relations to a Nearby Molecular Cloud}

\author{Y. Uchiyama}{
  address={Yale Center for Astronomy and Astrophysics, Yale University,
  P.O. Box 208121, \\ New Haven, CT 06520-8121, USA}
}

\author{F. A. Aharonian}{
  address={Max-Planck-Institut f\"{u}r Kernphysik, Postfach 103980,
        D-69029 Heidelberg, Germany}
}

\author{T. Takahashi}{
  address={Institute of Space and Astronautical Science, JAXA, 
        Sagamihara, Kanagawa 229-8510, Japan}
  ,altaddress={Department of Physics, University of Tokyo, 
  Bunkyo-ku, Tokyo 113-0033, Japan} 
}

\author{J. S. Hiraga}{
  address={Institute of Space and Astronautical Science, JAXA, 
        Sagamihara, Kanagawa 229-8510, Japan}
}

\author{Y.~Moriguchi}{
  address={Department of Astrophysics, Nagoya University, 
        Furo-cho, Chikusa-ku, Nagoya 464-8602, Japan}
}

\author{Y. Fukui}{
  address={Department of Astrophysics, Nagoya University, 
        Furo-cho, Chikusa-ku, Nagoya 464-8602, Japan}
}

\begin{abstract}
The recent X-ray and CO observations of  RX~J1713.7$-$3946 
show that a significant fraction of the nonthermal X-ray emission of this  
unique supernova remnant  associates,  in one way or another,  
with a molecular cloud interacting with the west part of the shell.
This adds a new puzzle  in the origin of  X-ray emission   
which cannot be easily explained within the standard model  
in accordance of which X-rays  are  result of synchrotron radiation 
of multi-TeV electrons accelerated by strong  shock waves. 
We explore an alternative origin of  the X-ray emission  assuming that it is 
produced  by secondary $e^{\pm}$ resulting from 
high energy hadronic interactions in the molecular gas.  
Such a scenario could explain in a quite natural way the apparent 
correlation  between the X-ray and CO morphologies. However,  
the TeV $\gamma$-ray emission recently reported by  H.E.S.S. 
significantly constrains the parameter space of this model. 
Namely, this mechanism cannot reproduce the bulk of 
the observed X-ray flux unless one postulates existence of a
{\it PeV cosmic-ray component} 
penetrating with an unusually hard spectrum into the dense cloud. 
\end{abstract}

\maketitle


\section{SNR RX~J1713.7$-$3946: Brief summary}

The SNR RX J1713.7$-$3946 exhibits remarkable nonthermal X-ray emission, 
with the largest total energy flux of 
$\simeq 5\times 10^{-10}\ \rm erg\ cm^{-2}\ s^{-1}$ (2--10 keV) 
amongst the shell-type SNRs. The diffuse X-ray emission, presumably of synchrotron 
origin, is distributed over the entire  SNR. It is   characterized by 
a power-law spectrum with an average photon index $\Gamma \simeq 2.3$ 
\cite{koyama97,slane99}.

The observations of this source with \emph{Chandra} 
\cite{uchiyama03, lazendic04}
revealed  that 
the bright  northwestern (NW) rim is comprised of 
a complex network of filaments and hotspots embedded 
in diffuse emission.
The X-ray spectra of these sub-components in the NW rim are well 
fitted by a power law with variation of the photon indices 
within  $\Gamma \simeq 2.1\mbox{--}2.5$
 \cite{uchiyama03}.
Recently, similar spatial and spectral characteristics have been 
found  in the southwestern (SW) rim in   
the \emph{XMM-Newton} data \cite{cassam04, hiraga04}.  
The applicability of the standard shock-acceleration 
(more specifically, diffusive shock acceleration) scenario 
to the synchrotron X-ray emission in RX~J1713.7$-$3946 
has been disputed based on the lack of indication of the expected 
(within the framework of this model) cutoff energy below  
10 keV \cite{uchiyama03}.

Originally it has been claimed that the molecular clouds 
found at a distance of 6 kpc are associated 
with this SNR \cite{slane99}. Therefore the distance to the 
SNR has been widely  adopted as 6 kpc. 
However,  the recent CO observations
made with the sensitive \emph{NANTEN} telescope have revealed 
a strong indication of  interaction of   
a molecular cloud located at a distance of 1 kpc
with RX~J1713.7$-$3946 \cite{fukui03}. This finding is  
further strengthened by an elaborate comparison between 
the X-ray and molecular emissions presented below and by 
the X-ray absorption features \cite{uchiyama04}.
The atomic and molecular hydrogen column density 
toward the line-of-sight of the  X-ray nebula 
becomes comparable to X-ray absorbing 
column density at $\simeq 1\ \rm kpc$ \cite{koo04}, which also supports 
the association with the nearby molecular cloud. 
The $\simeq 1\ \rm kpc$ distance to the supernova remnant 
makes  plausible
its association with the guest star A.D. 393 \cite{wang97}
(the remnant age of $t_{\rm age} \simeq 1600$ yr).

The CANGAROO collaboration reported  
the detection of TeV $\gamma$-ray emission from the direction 
of the NW rim \cite{muraishi00,enomoto02}.
The spectrum  was claimed to be quite steep
with  a  power-law photon index 
$\Gamma = 2.8\pm 0.2$ (0.4--8 TeV). It was 
argued that the steep spectrum  is inconsistent with the 
inverse Compton model,  but could be explained 
by $\pi^0$-decay $\gamma$-rays \cite{enomoto02}.
Most recently, the H.E.S.S. collaboration announced detection of  
TeV $\gamma$-rays \emph{distributed 
over the entire  X-ray nebula} with 
a hard  power-law spectrum $\Gamma = 2.2\pm 0.1$ (1--10 TeV) 
and flux $3.5\times 10^{-11}\ \rm erg\ cm^{-2}\ s^{-1}$ \cite{hess04}.
The fascinating $\gamma$-ray image detected by H.E.S.S. appears to be 
generally  similar to (although  not exactly same as) the X-ray image.

In this paper we  compare the spatial structures of the nonthermal X-ray nebula RX~J1713.7$-$3946 with the CO image of 
the interacting molecular cloud, 
and briefly discuss the origin of X-radiation which seems 
to be quite different from synchrotron X-ray emissions of 
other young shell-type SNRs \cite{vink04}, and   
explore the capability of secondary $e^{\pm}$, generated through hadronic 
collisions in the cloud, to explain the X-ray nebula.
Details will be presented elsewhere \cite{uchiyama04}.

\section{RX~J1713.7$-$3946: Comparison between Nonthermal X-ray and CO emissions}

\begin{figure}
  \includegraphics[height=.4\textheight]{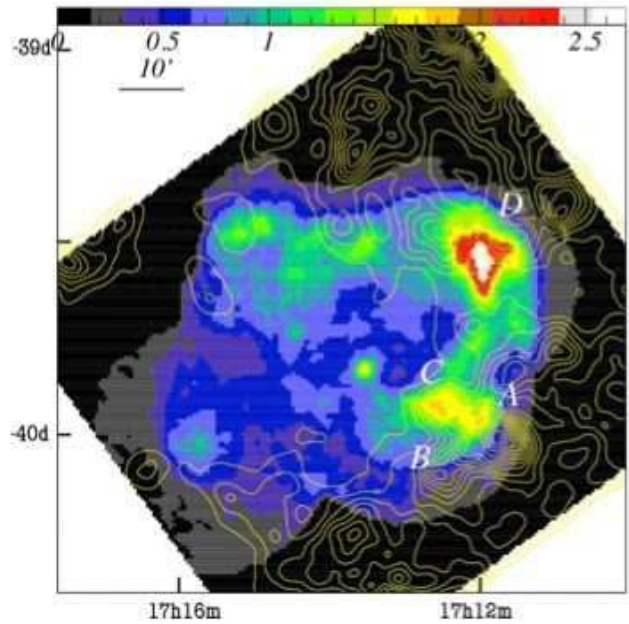}
  \caption{
  \emph{ASCA} X-ray image (1--5 keV) of RX~J1713.7$-$3946  
overlaid with CO($J$=1--0) map 
(contours) in the velocity range from $-11$ to $-3\ \rm km\ s^{-1}$.
The X-ray intensity scale is counts ks$^{-1}$ pixel$^{-1}$ 
with a pixel size of $\simeq 0.5\arcmin$.
The CO contours are every $2\ \rm K\ km\ s^{-1}$ in 
the range of  2--30 K km s$^{-1}$.
North is up and east is to the left.}
   \label{fig:asca_nanten}
\end{figure}

In Fig.~\ref{fig:asca_nanten}, the overall X-ray nebula 
is compared with the \emph{NANTEN} CO distribution
of the  nearby interacting molecular cloud.
The X-ray nebula shapes a large parallelogram 
($60\arcmin \times 70\arcmin$), 
with enhanced brightness in the northwest and southwest parts,   which 
we refer  as  ``NW rim''  and   ``SW rim'', respectively. 
These rims with an apparent width $\sim 10\arcmin$ have been extensively examined in Refs.~\cite{uchiyama03,hiraga04,lazendic04,cassam04}.
The bright NW and SW rims appear to be buried in the molecular cloud.
Four CO peaks (A, B, C, and D) are all located in the western part of 
the nebula. 
A direct correlation between the X-ray and CO emissions 
can be seen for peak C, 
from which the broad-line component of  CO  emission, 
a good indicator of SNR interactions \cite{white87}, 
has been detected \cite{fukui03}.
This   morphological comparison suggests that 
the dense molecular gas somehow plays a crucial  role for the 
formation of the bright nonthermal X-ray rims. 

\begin{figure}
  \includegraphics[width=.38\textheight]{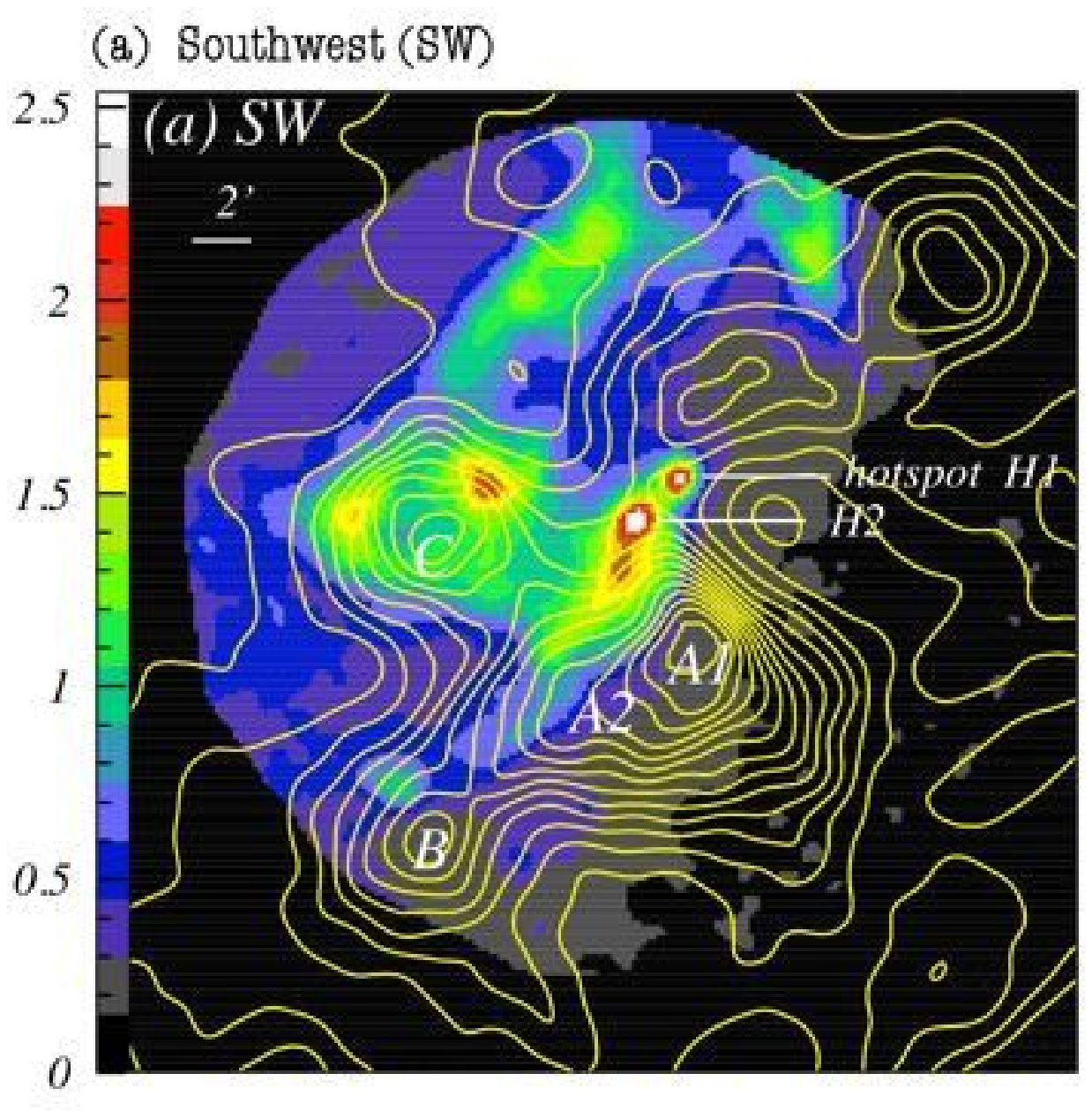}
  \includegraphics[width=.38\textheight]{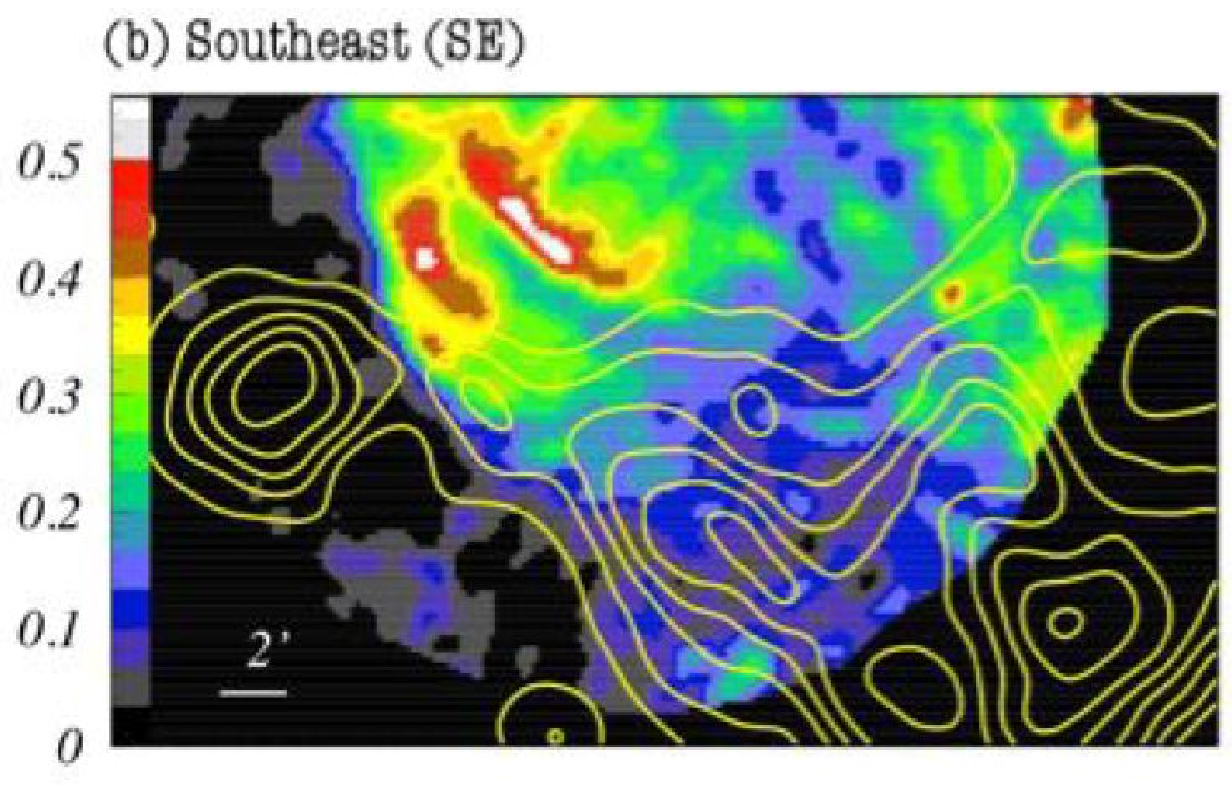}
     \caption{
\emph{XMM} MOS images (0.6--5 keV) 
accumulated in $10\arcsec \times 10\arcsec$ 
pixels with the same CO map used in Fig.~\ref{fig:asca_nanten} overlaid.
The X-ray intensity scale is counts ks$^{-1}$ pixel$^{-1}$.
North is up and east is to the left.
{\bf a)}
For the SW part. 
The CO contours are every $2\ \rm K\ km\ s^{-1}$ from 
2 to 30 K km s$^{-1}$.
{\bf b)}
For the SE part. 
The CO contours are from 1 to 8 K km s$^{-1}$
with an interval of $1\ \rm K\ km\ s^{-1}$.
}
   \label{fig:sw_image}
\end{figure}

A close-up X-ray image of the SW portion, where the major CO 
peaks are located, is displayed in Fig.~\ref{fig:sw_image}a 
together with  the CO($J$=1--0) intensity map. 
The \emph{XMM-Newton} image demonstrates that the
CO  peak C has 
excellent spatial match with the nonthermal X-ray emission.
Through spectral analysis, we found that the X-ray emission from 
peak C experiences additional soft X-ray absorption 
($\Delta N_{\rm H} \simeq 2\times 10^{21}\ \rm cm^{-2}$), 
confirming the association with the CO peak.
Also we found another bright X-ray feature that correlates well with 
the CO clumps, lying at the northeastern  
edge (a steep gradient of the CO line intensity)
of peak A (A1+A2). 
The X-ray spectra of these bright features have 
a photon index $\Gamma \simeq 2.3$, quite  typical for the west part of the rim.

In Fig.~\ref{fig:sw_image}b,
we also present the X-ray image of the southeastern (SE) sub-field, 
where the X-ray intensity is very low. 
The faint diffuse X-ray emission seems to  associate  with the molecular gas 
in the SE field.  Interestingly, 
while in the SW rim the {\em brightest} X-ray features emanate from the 
molecular cloud, in the SE part the {\rm faintest}  appear to originate 
from the cloud. 
The X-ray spectrum extending up to at least 5 keV suggests 
nonthermal origin, although the faintness of the X-ray emission 
from this region does not  unfortunately allow a detailed  spectral 
study. 

The results presented in Figs.~\ref{fig:sw_image}a,b  imply  that  
the nonthermal  X-ray emission 
emanates directly from dense molecular gas 
(though it cannot be excluded that 
a part of the emission originates in  hot tenuous plasma).
Usual synchrotron interpretation for the X-ray emission 
requires presense  of multi-TeV electrons in the molecular cloud.
Acceleration of multi-TeV electrons \emph{inside} the molecular cloud, 
however, is unlikely with conventional shock-acceleration models. 
Slow shock waves propagating inside the dense cloud 
($v \ll 1000\ \rm km\ s^{-1}$)
are not fast enough to accelerate electrons to very high energies;
the diffusive shock acceleration time scales as 
$t_{\rm acc} \propto v^{-2}$, where $v$ is the shock velocity. 

Yet \emph{external acceleration} scenarios also involve some difficulties. 
Suppose, for example,  that the 
fast supernova shock expanding into  the stellar wind bubble 
accelerates  multi-TeV electrons and injects  them 
into the molecular cloud resulting in  synchrotron X-ray emission.
The ultra-relativistic electrons are subject of  severe synchrotron losses 
as passing through  the magnetized cloud, with 
synchrotron cooling time $t_{\rm syn} \simeq 240 \ \rm yr$ 
for a moderate magnetic field of $B=30\,\mu$G. 
Unless the X-ray nebula is extremely young ($t_{\rm age}< t_{\rm syn}$), 
one should observe \emph{aging} effects of synchrotron-emitting electrons. 
Nevertheless, we measured \emph{hard} synchrotron X-ray spectra 
($\Gamma \simeq 2.3$) throughout the NW and SW rims, 
which seems to argue against external origins  
of the X-ray emitting electrons. 

The power-law spectra of synchrotron X-rays 
measured in this nebula suggest synchrotron cutoffs located 
at energies beyond  $\epsilon_0 \ga 10$ keV \cite{uchiyama03}.
On the other hand, within the standard framework of diffusive shock acceleration, 
the synchrotron cutoff should be located well below 10 keV.   

Therefore it is important to explore other possible (alternative) models 
 for interpretation of the spectral and spatial features of 
 the observed nonthermal X-ray emission. 
Below  we discuss the capability of a hadronic model  to explain 
a non-negligible fraction of X-rays by synchrotron emission of 
electrons from decays of secondary $\pi^\pm$ mesons 
produced in hadronic interactions.

\section{Secondary-electron synchrotron X-radiation}

Interactions of accelerated protons with the ambient gas lead 
to the production of $\pi$-mesons  with  comparable fractions of the energy 
of primary protons transferred to  the final products  of $\pi$-meson 
decays---$\gamma$-rays, electrons and neutrinos.  
While $\gamma$-rays and neutrinos freely escape their production regions
and can be observed directly, the secondary electrons lose their energy 
through synchrotron radiation and inverse Compton scattering. 
In presence of sufficiently large magnetic field, e.g.\ close to 
$100 \ \mu \rm G$ or more,
which is needed  for acceleration of protons to PeV energies, 
the bulk of the electron energy is released in the form of hard X-rays.  
The production of X-rays by secondary electrons is an attractive possibility which to a large extent is free of  the problems associated 
with the synchrotron radiation by {\it directly accelerated electrons}.
In particular, this model allows diffuse synchrotron X-ray emission 
throughout the remnant (because the protons can propagate without significant losses), as well as allows extension of X-ray spectra 
well above 10 keV. 

One may expect that 
\emph{directly accelerated} electrons 
are dominant contributers to synchrotron X-ray emission, 
rather than the secondary electrons.  Actually, indeed the  
primary synchrotron X-radiation can exceed the contribution by the 
secondary electrons  by more than an order of magnitude 
for shell-type SNRs with low  ambient density, 
e.g.\ $n \sim 1\ \rm cm^{-3}$ and 
a proton-to-electron ratio $\sim 100$.
However, since the cease of particle injection quickly 
results in  a dramatic drop of the X-ray flux by primary electrons 
on a timescale $t_{\rm syn} \approx 100 \ \rm yr$,  while the  
X-ray flux by secondaries remains unchanged on much longer 
timescale $\tau_{\rm pp} \ga 10^5$ yr,  in the case of a  
``dead'' accelerator  (e.g.\ due to the crash of the shell with 
a dense cloud) we may expect X-radiation only from 
secondary electrons. Moreover, in dense environments, 
like molecular clouds interacting with the SNR shell, 
the synchrotron X-radiation may dominate even in the case of 
an active accelerator.   

Finally,  the long-lived ultrarelativistic protons can propagate 
throughout the dense gas 
generating secondary $e^{\pm}$ via 
the production and subsequent decay of charged pions.
Unlike directly accelerated electrons which are expected to be 
concentrated in the proximity of their acceleration sites, 
the secondary electrons can be easily distributed over 
larger volume, resulting in  extended synchrotron radiation as observed. 

\begin{figure}
  \includegraphics[width=.7\textwidth]{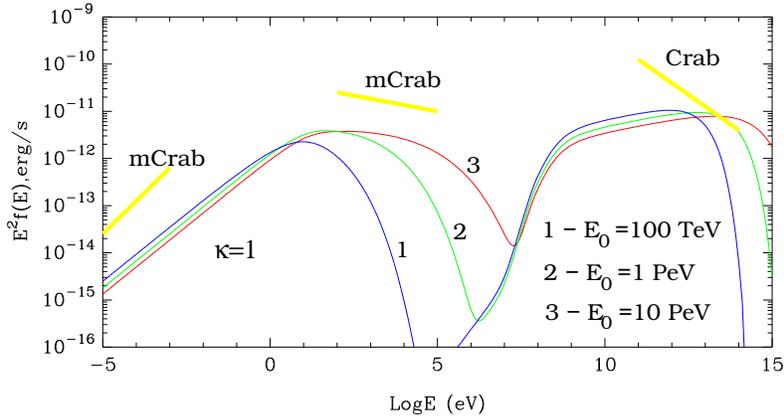}
            \caption{The broadband SEDs of  radiation 
                (secondary $e^{\pm}$ synchrotron and 
            $\pi^0$-decay gamma-rays) 
initiated by \emph{p--p} interactions. It is asumed that protons are 
injected continuously into a dense region with linear size 
$R=1\ \rm pc$, gas density $n = 100\ \rm cm^{-3}$,
and magnetic field $B=100\ \mu\rm G$. The injection spectrum of 
protons is a power law with spectral index $s=1.9$ and 
exponential cutoff $E_0=0.1$ PeV, 1 PeV, and 10 PeV. 
The source age 1000 yr. The proton injection power 
$L_{\rm p} = 3\times 10^{37}\ \rm erg\ s^{-1}$ 
(total injected energy $W_{\rm p} \simeq  1\times 10^{48}\ \rm erg$). 
The escape time of protons in the source is set assuming Bohm diffusion.}
\label{fig:ppSED}
\end{figure}

In Fig.~\ref{fig:ppSED}, 
we show a characteristic example of  broad-band SEDs of 
nonthermal radiation resulting from \emph{pp} interactions, assuming that 
protons are injected into the production region with a (number) 
spectrum $Q\propto E^{-1.9} e^{-E/E_0}$. 
It is seen that for the  sufficiently high cutoff energy in the spectrum 
of protons ($E_0 \geq 1 \ \rm PeV$), the secondary synchrotron X-ray flux  becomes comparable to that of $\pi^0$-decay $\gamma$-rays. 
Since the $\pi^0$-decay $\gamma$-rays are tightly coupled 
in such a way with the synchrotron radiation of secondary electrons, 
the TeV $\gamma$-rays flux detected by H.E.S.S. \cite{hess04}
constrains the level of the X-ray flux contributed by secondary electrons.  
Note that the  $\pi^0$ components shown  in Fig.~\ref{fig:ppSED}
correspond to the H.E.S.S. flux from the NW part.
Since the accompanying X-ray flux by secondary-electrons falls well  below 
(by an order of magnitude) the observed X-ray flux 
 (at several mCrab level), one may conclude that the contribution 
by secondary electrons cannot exceed 10 percent of the observed X-ray flux. 

However, the relative contribution of X-rays can be significantly increased 
without violating the H.E.S.S. flux 
if one assumes very hard proton spectrum, e.g.\ $Q\propto E^{-1.5}$,
extending to PeV energies.  
It should be noted that such a hard component of cosmic rays
is not excluded  by the nonlinear
shock acceleration theory \cite{malkov01}.
Also, very hard proton spectra can be expected due to
propagation effect.  For example, it is possible, at least in principle, that 
due to  very slow diffusion only $\geq 1 \ \rm PeV$ protons can 
arrive, for the limited age of the accelerator, 
at the dense regions of the clouds.
In this case formally we can expect an arbitrary hard proton spectrum which in the regions of enhanced density may initiate secondary radiation with  unusual SEDs---with distinct peaks at the hard X-ray and ultrahigh energy 
$\gamma$-ray domains.





\bibliographystyle{aipproc}   

\bibliography{yu_hd04}

\IfFileExists{\jobname.bbl}{}
 {\typeout{}
  \typeout{******************************************}
  \typeout{** Please run "bibtex \jobname" to optain}
  \typeout{** the bibliography and then re-run LaTeX}
  \typeout{** twice to fix the references!}
  \typeout{******************************************}
  \typeout{}
 }

\end{document}